\pdfoutput=1

\documentclass[journal]{IEEEtran}
\usepackage{cite}

\usepackage[pdftex]{graphicx}

\usepackage{amsmath}

\usepackage[small]{subfigure}
\usepackage{multirow}

\usepackage[pscoord]{eso-pic}% The zero point of the coordinate systemis the lower left corner of the page (the default).
\usepackage{textcomp}
\usepackage{hyperref}
\newcommand{\placetextbox}[3]{% \placetextbox{<horizontal pos>}{<vertical pos>}{<stuff>}
  \setbox0=\hbox{#3}% Put <stuff> in a box
  \AddToShipoutPictureFG*{% Add <stuff> to current page foreground
    \put(\LenToUnit{#1\paperwidth},\LenToUnit{#2\paperheight}){\vtop{{\null}\makebox[0pt][c]{#3}}}%
  }%
}%

\begin{document}

\placetextbox{0.5}{0.995}{\normalfont \small \textcopyright 2020 IEEE. Personal use of this material is permitted. Permission from IEEE must be obtained for all other uses, in any current}

\placetextbox{0.5}{0.985}{\normalfont \small or future media, including reprinting/republishing this material for advertising or promotional purposes, creating new collective}

\placetextbox{0.5}{0.975}{\normalfont \small works, for resale or redistribution to servers or lists, or reuse of any copyrighted component of this work in other works}

\placetextbox{0.5}{0.96}{\normalfont \small Author accepted manuscript, published in ``IEEE Transactions on Circuits and Systems II: Express Briefs, 2018, 65(12): 2067--2071''.}

\placetextbox{0.5}{0.945}{\normalfont DOI: \url{https://doi.org/10.1109/TCSII.2018.2821920}.}%

\title{A Multi-Objective Approach for Post-Nonlinear Source Separation and its Application to Ion-Selective Electrodes}

\author{Guilherme~D.~Pelegrina,~\IEEEmembership{Member,~IEEE,}
        Leonardo~T.~Duarte,~\IEEEmembership{Senior~Member,~IEEE}% <-this % stops a space
\thanks{This work was supported by FAPESP (Process n. 2014/27108-9 and 2015/16325-1) and CNPq (Process n. 311786/2014-6).}% <-this % stops a space
\thanks{G. D. Pelegrina and L. T. Duarte are with the School of Applied Sciences, University of Campinas (UNICAMP), Limeira, Brazil (e-mail:{guilherme.pelegrina,leonardo.duarte}@fca.unicamp.br).}}

\markboth{}%
{Shell \MakeLowercase{\textit{et al.}}: Bare Demo of IEEEtran.cls for IEEE Journals}

\maketitle

\begin{abstract}
Blind source separation (BSS) methods have been applied to deal with the lack of selectivity of ion-selective electrodes (ISE). In this paper, differently from the standard BSS solutions, which are based on the optimization of a mono-objective cost function associated with a given property of the sought signals, we introduce a novel approach by relying on multi-objective optimization. Numerical experiments with actual data attested that our proposal allows the incorporation of additional information on the interference model and also provides the user a set of solutions from which he/she can select a proper one according to his/her prior knowledge on the problem.
\end{abstract}


\IEEEpeerreviewmaketitle

\section{Introduction}
\label{sec:intro}
\IEEEPARstart{A}{n} ion-selective electrode (ISE) is a device used to measure the ionic activity of a target ion in an aqueous solution~\cite{Grundler2007}. Due to the interference of ions other than the target one, the ISE response may contain a mixture of different ionic activities. An interesting approach to tackle the interference problem is to consider an acquisition step based on more than one ISE followed by a multivariate signal processing step whose aim is to separate each of the ionic activities. With respect to this second step, blind source separation (BSS)~\cite{Comon2010} methods have been providing an interesting solution. Among other advantages, they are able to simplify the calibration step that is often required in the signal processing step~\cite{Duarte2009,Duarte2014}.

In short, a BSS method aims at recovering a set of source signals (or simply sources) from a set of mixtures of theses sources. The term \textit{blind} is used since the separation is carried out by only considering the mixtures and by only assuming a parametric information on the mixing process. BSS has been successfully applied in a great number of domains, from biomedical signal analysis~\cite{Comon2010} to microchip fabrication~\cite{2010SchacPL}.  

In the context of ISE analysis, the sources correspond to the different ionic activities and the mixtures are the responses acquired by the electrodes. In addition, there is a challenging aspect related to the fact that interference in ISE is clearly a nonlinear phenomenon, and, thus, nonlinear BSS methods must be considered in this case~\cite{2015DevilD,Comon2010,Woo2005,Solazzi2004,Sun2009}. In particular, the nonlinear mixing process that takes place at ISEs can be well described by the class of post-nonlinear (PNL) models~\cite{Taleb1999}.

Several BSS techniques have been developed to deal with the interference problem in ISEs, e.g. Bayesian approach~\cite{Duarte2009} and independent component analysis (ICA) algorithm~\cite{Duarte2014}. Furthermore, in~\cite{Ando2015}, the authors proposed a BSS method considering a quadratic mixing model. Classically, the task of recovering the ionic activities can be accomplished by formulating a mono-objective optimization problem in which the cost function is related to a given property of the original sources. For instance, in ICA~\cite{Comon2010,Yang2008}, one assumes that the sources are statistically independent and, thus, the resulting optimization problem boils down to the minimization of a measure of statistical dependence of the retrieved sources.

In this paper, differently from the standard mono-objective-based BSS methods, we introduce a novel approach that relies on a multi-objective formalism~\cite{Miettinen1999}. A first motivation for our idea is that, by considering a multi-objective approach, it becomes possible to incorporate different prior information into the solution of the problem. This can be done, for instance, by assigning an objective function for modeling a known property of the sources and another one to take into account information provided by the physical-chemical modeling of the interference problem. Another advantage of our proposal is that it provides a set of non-dominated solutions, from which the user can choose one signal based on his/her prior knowledge on the problem. For instance, the user may select sources according to subjective priors that are typical in chemistry (e.g., smoothness of observed signals). 

The paper is organized as follows. We first provide a brief description of the BSS problem associated with the mixing process related to ISEs. In Section~\ref{sec:proposed_approach}, we present the proposed multi-objective BSS method. In Section~\ref{sec:experimental_results}, we provide numerical experiments considering actual data acquired by two ISEs. Finally, our conclusions are discussed in Section~\ref{sec:conclusion}.

\section{Problem statement}
\label{sec:problem_statement}

In an ISE, the parametric information of the mixing process can be obtained by the Nicolsky-Eisenman (NE) equation~\cite{Grundler2007}. Indeed, according to this equation, the signal acquired by the $i$-th ISE within an array, that is, the $i$-th mixture, is given by
\begin{equation}
\label{intro:pnl}
x_i(t)=e_i+d_i\log_{10}{\left(s_i(t)+\sum_{j=1,j \neq i}^N{a_{ij}s_j(t)}\right)},
\end{equation}
where $e_i$ and $d_i$ are constants, $s_i(t)$ and $s_j(t)$ represent the activities of the target and the interfering ions, respectively, i.e $s_k(t)$ correspond to the $k$-th source. The parameters $a_{ij}$ are the selective coefficients and $N$ is the number of sources (ionic activities to be estimated). Therefore, the task of BSS in this case is to estimate the set of signals $\mathbf{s}(t) = [s_1(t), s_2(t), \ldots, s_N(t)]$ from the set of mixtures  $\mathbf{x}(t) = [x_1(t), x_2(t), \ldots, x_M(t)]$. $e_i$, $d_i$ and the coefficients $a_{ij}$ are also unknown and, thus, must be adjusted. In the experiment conducted in this work, the number of mixtures (that is, the number of ISEs) and of sources are the same: $M=N$\footnote{Our proposal can be extended to the case $M>N$, as in the linear case~\cite{Comon2010}.}.

In view of the scaling ambiguity typical of BSS methods~\cite{Comon2010}, the parameters $e_i$ cannot be blindly estimated. Therefore, by setting these parameters zero, for instance, the model is still identifiable\footnote{Note that if the separating system $\mathbf{y}(t)=\mathbf{W}\left(10^{(\mathbf{x}(t)-\mathbf{e}^*) \circ \frac{1}{\mathbf{d}^*}}\right)$ were considered (in contrast to system of Eq.~(\ref{intro:pnlinv})), one would end up with $\mathbf{y}(t)=\mathbf{W}\left(10^{\mathbf{x}(t) \circ \frac{1}{\mathbf{d}^*}}\right) \left(10^{-\left(\mathbf{e}^* \circ \frac{1}{\mathbf{d}^*}\right)}\right)$, which means that $\mathbf{e}^*$ would only change the scale of the retrieved sources. Since there is a scaling ambiguity in BSS, it is thus not possible to estimate $\mathbf{e}^*$. The price to be paid here is that the retrieved sources may differ from the original ones by a constant gain, and, thus, some few calibration points are required in the end of the separation process~\cite{Duarte2009}.}, which leads us to the following mixing model expressed in vector notation:
\begin{equation}
\label{intro:pnlchem}
\mathbf{x}(t)=\mathbf{d} \circ \log_{10}{\left(\mathbf{A}\mathbf{s}(t)\right)},
\end{equation}
where $\mathbf{d}=[d_1, d_2, \ldots,d_N]$, $\mathbf{A}=\left[a_{ij}\right]$ and the operator $\circ$ represents the Hadamard product.

Given the mixing process expressed by Equation~\eqref{intro:pnlchem}, BSS can be performed by adjusting a set of parameters $\mathbf{d}^*$ and a separating matrix $\mathbf{W}$ in order to find a set of signals $\mathbf{y}(t) =[y_1(t), y_2(t), \ldots, y_N(t)]$ given by 
\begin{equation}
\label{intro:pnlinv}
\mathbf{y}(t)=\mathbf{W}\left(10^{\mathbf{x}(t) \circ \frac{1}{\mathbf{d}^*}}\right)
\end{equation}
that are as close as possible to $\mathbf{s}(t)$.

\section{A multi-objective blind source separation method for ion-selective electrodes}
\label{sec:proposed_approach}

\subsection{Basic concepts in multi-objective optimization}

In mono-objective optimization, the solution is achieved by optimizing a single cost function. In the context of Equation~\eqref{intro:pnlinv}, this approach can be formulated as%\footnote{We considered a minimization problem in this formulation, however, it can be transformed to maximization by proceeding the simple transformation $\min \, J(\cdot)=\max \, -J(\cdot)$.}
\begin{equation}
\label{eq:mono}
\underset{\mathbf{W},\mathbf{d}^*}{\min} \, \, J(\mathbf{W},\mathbf{d}^*),
\end{equation}
where $J(\cdot)$ is a separation criterion which is related to a given property of the sources. In ICA, $J(\cdot)$ can be defined as the mutual information between the estimated sources. 

Conversely, when additional prior information is available, it becomes possible to formulate the problem by considering two or more optimization criteria. In the problem addressed in this paper, this could be done by defining cost functions $J_i(\mathbf{W},\mathbf{d}^*)$ that would be associated with each information at hand. A first idea is thus to combine the set of cost functions into a single one, by adding weights according to their importance in the separation problem. Another alternative is to define a given criterion as the single cost function and to set the remaining criteria as constraints, each one with a specific lower/upper limit. However, these approaches are sensitive to the predefined parameters (weights and limits, respectively, which are often unknown), and, thus, different parameters may lead to different solutions.

In the present paper, instead of formulating the problem as a mono-objective one, we consider a multi-objective optimization approach which can be expressed as follows:
\begin{equation}
\label{eq:multi}
\underset{\mathbf{W},\mathbf{d}^*}{\min} \, \, \mathbf{J}(\mathbf{W},\mathbf{d}^*)=\underset{\mathbf{W},\mathbf{d}^*}{\min} \, \, [J_1(\mathbf{W},\mathbf{d}^*), \ldots, J_K(\mathbf{W},\mathbf{d}^*)],
\end{equation}
where $\mathbf{J}(\cdot)$ represents the set of $K$ optimization criteria. An interesting aspect of multi-objective problems arises when the cost functions are conflicting. In these situations, instead of converging to a single solution to the problem, which is the case in mono-objective problems such as~\eqref{eq:mono}, one achieves a set of non-dominated solutions\footnote{Since we applied an evolutionary algorithm in our experiments, which does not guarantee optimality, we adopt the term ``non-dominated solutions'' instead of ``Pareto optimal solutions''.}~\cite{Miettinen1999}. In the problem expressed in~(\ref{eq:multi}), these non-dominated solutions represent a set of possible estimations for the ionic activities (sources). 

The non-dominated set is determined based on the concept of dominance. A solution $\mathbf{p}$ dominates another solution $\mathbf{q}$ if $\mathbf{p}$ is as good as $\mathbf{q}$ in all objectives and $\mathbf{p}$ is strictly better than $\mathbf{q}$ in at least one objective. Therefore, the non-dominated set comprises the solutions that are not dominated by any other solution~\cite{Deb2001}.

\subsection{Problem modeling}

The multi-objective approach proposed in this paper comprises two optimization criteria. The first one is often employed to perform source separation and it is built upon the assumptions that the sources have a temporal structure and are mutually uncorrelated. Such an approach has led to different BSS methods including the SOBI algorithm~\cite{Belouchrani1997,Liu2006}. When the sources are uncorrelated and modeled as stochastic processes, the covariance matrices for different delays $r$, denoted by $C_{\mathbf{s}\mathbf{s}}(r)$, and whose element $ij$ is given by $C_{\mathbf{s}\mathbf{s}_{ij}}(r) = E\{s_i(t)s_j(t-r)\}$, are diagonal matrices, since $E\{s_i(t)s_j(t-r)\} =0$ for all $r$ and $i\neq j$. However, due to the action of the mixing process, the covariance matrices of the mixtures, $C_{\mathbf{x}\mathbf{x}}(r)$, are not diagonal anymore. Indeed, in the case of linear mixtures, it asserts that $C_{\mathbf{x}\mathbf{x}}(r)=\mathbf{A}C_{\mathbf{s}\mathbf{s}}(r)\mathbf{A}^T$.

Since the covariance matrices $C_{\mathbf{x}\mathbf{x}}(r)$ are not diagonal ones, a natural idea to estimate the parameters of the separating system is to minimize a cost function related to the off-diagonal terms of the covariance matrices (for different delays $r$) of the retrieved sources, $C_{\mathbf{y}\mathbf{y}}(r)$~\cite{Comon2010}. Such an idea results in the following joint-diagonalization-based criterion:
\begin{equation}
\label{eq:sobi}
\sum_{r=0}^{R}{\left( \sum_{i,j, \, \, 1\leq i \neq j \leq N}^N \left(C_{\mathbf{y}\mathbf{y}_{ij}}(r)\right)^2 \right)}
\end{equation}
where $R$ is the number of delays and $C_{\mathbf{y}\mathbf{y}_{ij}}(r)$ denotes the element $ij$  of $C_{\mathbf{y}\mathbf{y}}(r)=E\{\mathbf{y}(t)\mathbf{y}(t-r)\}=E\{\mathbf{W}\left(10^{\mathbf{x}(t) \circ \frac{1}{\mathbf{d}^*}}\right)\mathbf{W}\left(10^{\mathbf{x}(t-r) \circ \frac{1}{\mathbf{d}^*}}\right)\}$. Note that $C_{\mathbf{y}\mathbf{y}}(r)$ is a function of $\mathbf{d}^*$ and $\mathbf{W}$. In this paper,~\eqref{eq:sobi} will be refered to as the SOBI criterion.

The separation criterion expressed in~\eqref{eq:sobi} relies on assumptions made on the sources but it does not take into account possible information on the mixing process. For instance, according to the NE equation, the slope parameters $d_i$ are related to physical constants and take 59 mV for room temperature and for monovalent ions. Therefore, even if deviations from this theoretical value can be observed in practice due to sensor aging (among other reasons), the Nernstian slope of 59 mV per decade can be considered as reference in the adaptation of $d_i^*$. This can be done, for instance, by considering an optimization criterion based on the mean squared error between $d_i^*$ and the reference values, that is, $\sum_{i=1}^N{\left(d_{i}^*-59\right)^2}$. This function will be refered to as the similarity with Nernstian slope criterion.

By considering the two criteria described above, the proposed BSS approach tackles the following multi-objective optimization problem\footnote{We considered $R=3$ delays, which seems enough to model the temporal aspect of chemical sources. It is worth mentioning that we performed experiments with $R=1$, $R=2$ and $R=4$, which provide similar results.}:
\begin{equation}
\label{met:moo}
\underset{\mathbf{d}^*}{\min}\, \, \left[\sum_{i=1}^N{\left(d_{i}^*-59\right)^2}, \, \sum_{r=0}^3{\left( \sum_{i,j, \, \, 1\leq i \neq j \leq N}^N \left(C_{\mathbf{y}\mathbf{y}_{ij}}(r)\right)^2 \right)}\right].
\end{equation}
As will be clarified in the sequel, the adjustment of the separating matrix $\mathbf{W}$ will be carried out in an implicit fashion by means of the SOBI algorithm~\cite{Belouchrani1997}, which explains why the formulated multi-objective problem depends explicitly only on the parameters $\mathbf{d}^*$. The SOBI algorithm performs the diagonalization of the covariance matrices $C_{\mathbf{y}\mathbf{y}}(r)$ through a Jacobi-like in which the minimization of the off-diagonal terms is conducted by successive Givens rotations (see~\cite{Belouchrani1997}).     

\subsection{Resolution strategy}

There are several techniques used to deal with multi-objective problems, such as the traditional ones~\cite{Miettinen1999} and the evolutionary algorithms~\cite{Deb2001}. The traditional techniques, e.g the weighting method and $\epsilon$-constraint, are less expensive compared to evolutionary algorithms. However, some prior information or initial analysis about the cost functions are required in this approach. For instance, a convexity analysis is required in the weighting method (in order to guarantee that all non-dominated solutions may be found) and an appropriate choice of the constraint bounds is needed in the $\epsilon$-constraint (in order to avoid infeasible solutions)~\cite{Miettinen1999}.

Despite the fact that the evolutionary algorithms are more computationally expensive compared to the aforementioned ones, they do not require further analysis about the problem. In view of this benefit, we consider the evolutionary algorithm called SPEA2~\cite{Zitzler2001} to tackle the multi-objetive problem of~\eqref{met:moo}. The main aspects of this algorithm are comprised in the following steps (for further details, see~\cite{Zitzler2001}):

\begin{enumerate}
\item \textbf{Initialization}: The first step comprises the random generation of the initial population $\mathbf{Z}=[\mathbf{d}^{1*}, \mathbf{d}^{2*}, \ldots, \mathbf{d}^{L*}]$, i.e. a set of $L$ candidates for the parameter $\mathbf{d}^*$ (which solves the multi-objective problem). An empty external set $\mathbf{\tilde{Z}}$ is also created to store the selected $\tilde{L}$ best candidates ($\mathbf{\tilde{d}}^{1*}, \mathbf{\tilde{d}}^{2*}, \ldots, \mathbf{\tilde{d}}^{\tilde{L}*}$) after each iteration.
%\item \textbf{Fitness assignment}: For each candidate in $\mathbf{Z}$ and $\mathbf{\tilde{Z}}$, we first calculate the cost functions (criteria) values (according to~\eqref{met:moo}). These values are used to analyze the dominance relation between all the candidates and, therefore, to determine a quality measure, known as fitness, for each one.
\item \textbf{Fitness assignment}: For each candidate $\mathbf{d}^{l*} \in \mathbf{Z}$ and $\mathbf{\tilde{d}}^{\tilde{l}*} \in \mathbf{\tilde{Z}}$, we first calculate the cost functions (criteria) values (according to~\eqref{met:moo}). Based on these values, we derive the total number of candidates that dominates each $\mathbf{d}^{l*}$ and $\mathbf{\tilde{d}}^{\tilde{l}*}$. This number\footnote{We may remark that it is equal to zero for non-dominated solutions}, combined with a density measure (which exploits the diversity of the solutions in the non-dominated set~\cite{Zitzler2001}), is used to determine a quality measure, known as fitness, for each candidate. 
\item \textbf{Selection}: The $\tilde{L}$ best candidates are selected according to their fitness (smaller is better), in order to update the external set $\mathbf{\tilde{Z}}$.
\item \textbf{Termination}: If a stopping criterion is satisfied (e.g. a maximum number $G$ of iterations) the algorithm stops and the non-dominated solutions (set of $\tilde{L}$ candidates for the parameter $\mathbf{d}^*$)  are represented by the ones stored in the external set $\mathbf{\tilde{Z}}$. Otherwise, we go to Step 5.
\item \textbf{Variation set selection}: The $L$ candidates that will be submitted to the variation step are selected via a binary tournament selection with replacement performed in $\mathbf{\tilde{Z}}$.
\item \textbf{Variation}: One here applies evolutionary operators of crossover (recombination of parameters from two different solutions) and mutation (alteration of one or more parameters in a single solution) to the variation set. The new population (with size $L$) generated by these operators ($\alpha\%$ by crossover and $(1-\alpha)\%$ by mutation) may provide better solutions when compared to the previous one. After performing the variation procedure, the iteration ends and the algorithm restart in Step 2.
\end{enumerate}

\begin{figure}[ht]
\centering
\includegraphics[height=3.50cm]{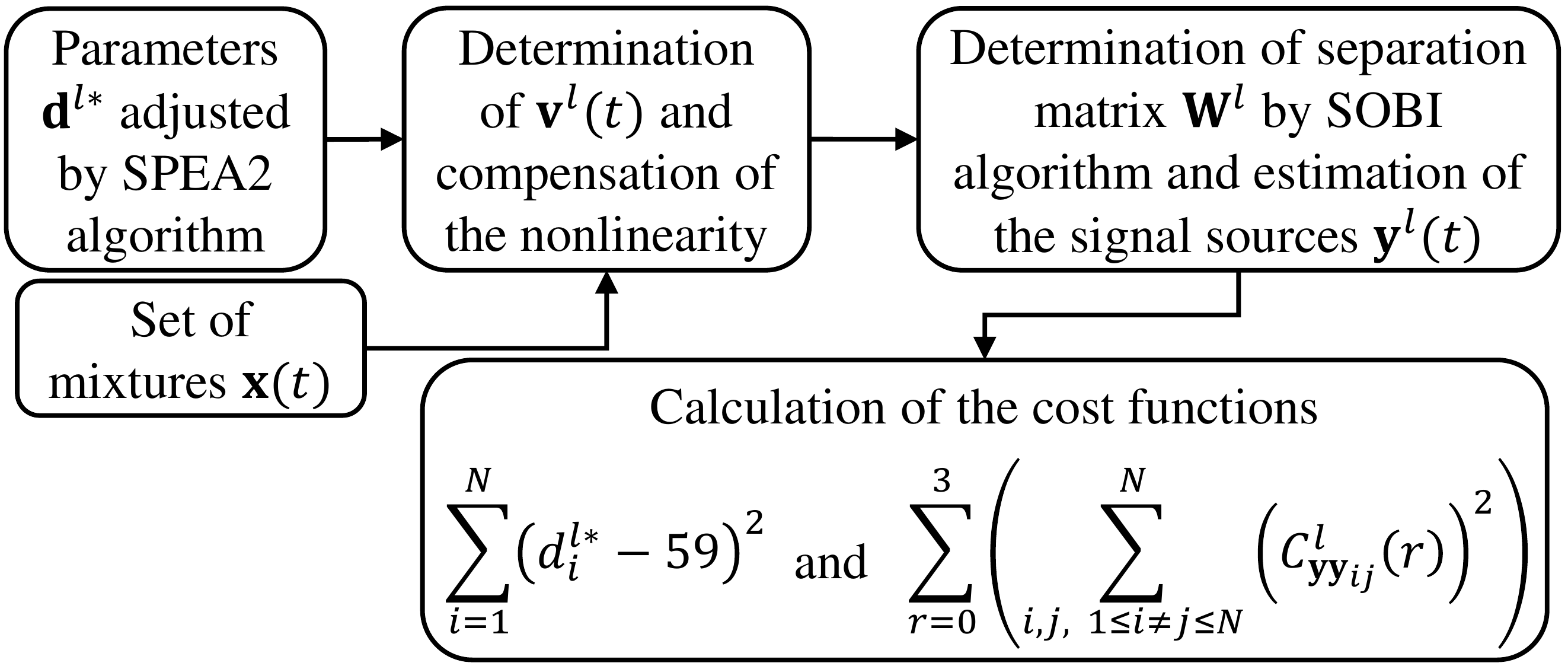}
\caption{A memetic strategy used in fitness assignement step to calculate the cost functions~\eqref{met:moo}.}
\label{fig:esqmem}
\end{figure}

As mentioned before, only the parameters $d_i^*$ are explicitly taken into account in the SPEA2 algorithm. Therefore, in the variation step, we perform adjustments only on $\mathbf{d}^{l*}$ (and $\mathbf{\tilde{d}}^{\tilde{l}*}$). In order to obtain the separating matrix $\mathbf{W}$ used to calculate the cost functions in the fitness assignment step ($C_{\mathbf{y}\mathbf{y}}(r)$ dependes on $\mathbf{W}$), we consider a memetic strategy~\cite{Dias2007}, which simplifies the nonlinear BSS problem expressed in~\eqref{intro:pnlinv}. In this strategy, given $\mathbf{d}^{l*}$ adjusted in every iteration of SPEA2, $\mathbf{W}^l$ is implicitly determined by an execution of the SOBI algorithm. Therefore, given $\mathbf{d}^{l*}$, we firstly tackle the effect of the nonlinearity such that $\mathbf{v}^l(t)=10^{\mathbf{x}(t) \circ \frac{1}{\mathbf{d}^{l*}}}$ and, after determining $\mathbf{W}^l$ by the SOBI algorithm, it becomes possible to tackle the linear transformation, as follows $\mathbf{y}^l(t)=\mathbf{W}^l\mathbf{v}^l(t)$. Figure~\ref{fig:esqmem} illustrates the the memetic strategy used to evaluate the cost functions in fitness assignment step for candidate $\mathbf{d}^{l*}$. Finally, it is worth recalling that the solution of a multi-objective problem is a set of non-dominated solutions~\cite{Miettinen1999}.

%As mentioned before, we consider a memetic strategy~\cite{Dias2007} to simplify the nonlinear BSS problem expressed in~\eqref{intro:pnlinv}. In this strategy, only the parameters $d_i^*$ are explicitly taken into account in the SPEA2 algorithm, since, for every iteration of the method, the separating matrix $\mathbf{W}$ is implicitly adjusted by an execution of the SOBI algorithm. Therefore, given the candidate $l$ in the population and the parameters $d_i^{l*}$ adjusted by the SPEA2 algorithm, we firstly tackle the effect of the nonlinearity such that $\mathbf{v}^l(t)=10^{\mathbf{x}(t) \circ \frac{1}{\mathbf{d}^{l*}}}$ and, after determining $\mathbf{W}^l$ by the SOBI algorithm, it becomes possible to tackle the linear transformation, as follows $\mathbf{y}^l(t)=\mathbf{W}^l\mathbf{v}^l(t)$. Figure~\ref{fig:esqmem} illustrates the evaluation of the cost functions using the memetic strategy for candidate $l$. These cost functions are used in the analysis of the dominance between candidates in order to determined their fitness. Finally, it is worth recalling that the solution of a multi-objective problem is a set of non-dominated solutions~\cite{Miettinen1999}.

\section{Experimental results}
\label{sec:experimental_results}

In order to assess our proposal, we consider an actual dataset which was acquired by two ISEs tailored to estimate the activities of the ions sodium (Na$^+$) and potassium (K$^+$), both measured in mV (see~\cite{Duarte2014}). The number of acquired samples was $41$. In the context of chemical analysis, one often must deal with problems in which the numbers of electrodes and samples are reduced.  

Since the dataset was acquired in a controlled experiment, the original sources are known and can be used as benchmark. In that spirit, we consider, as performance index, the signal-to-interference ratio (SIR, in dB), given by $SIR_i=10 \log_{10} \left(E\left\{s_i^2\right\}\left(E\left\{(s_i-\hat{y}_i)^2\right\}\right)^{-1}\right)$, where $\hat{y}_i$ is the $i$-th retrieved source after scaling correction.

\begin{figure}[ht]
\centering 
\includegraphics[height=2.30cm]{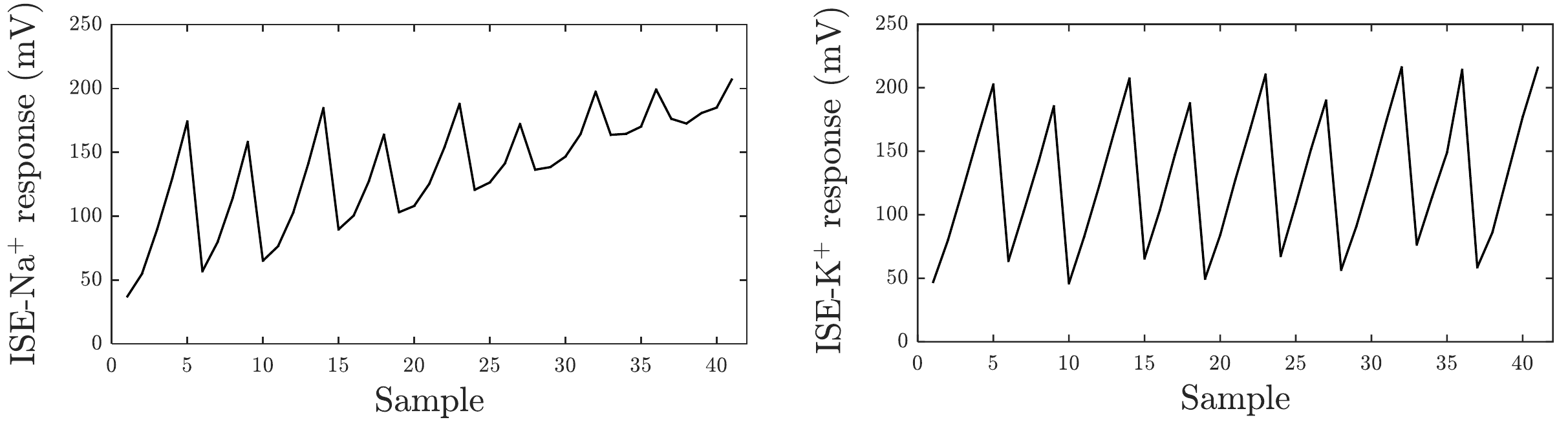}
\caption{ISE responses.}
\label{fig:mixt}
\end{figure}

\begin{figure}[ht]
\centering
\includegraphics[height=4.1cm]{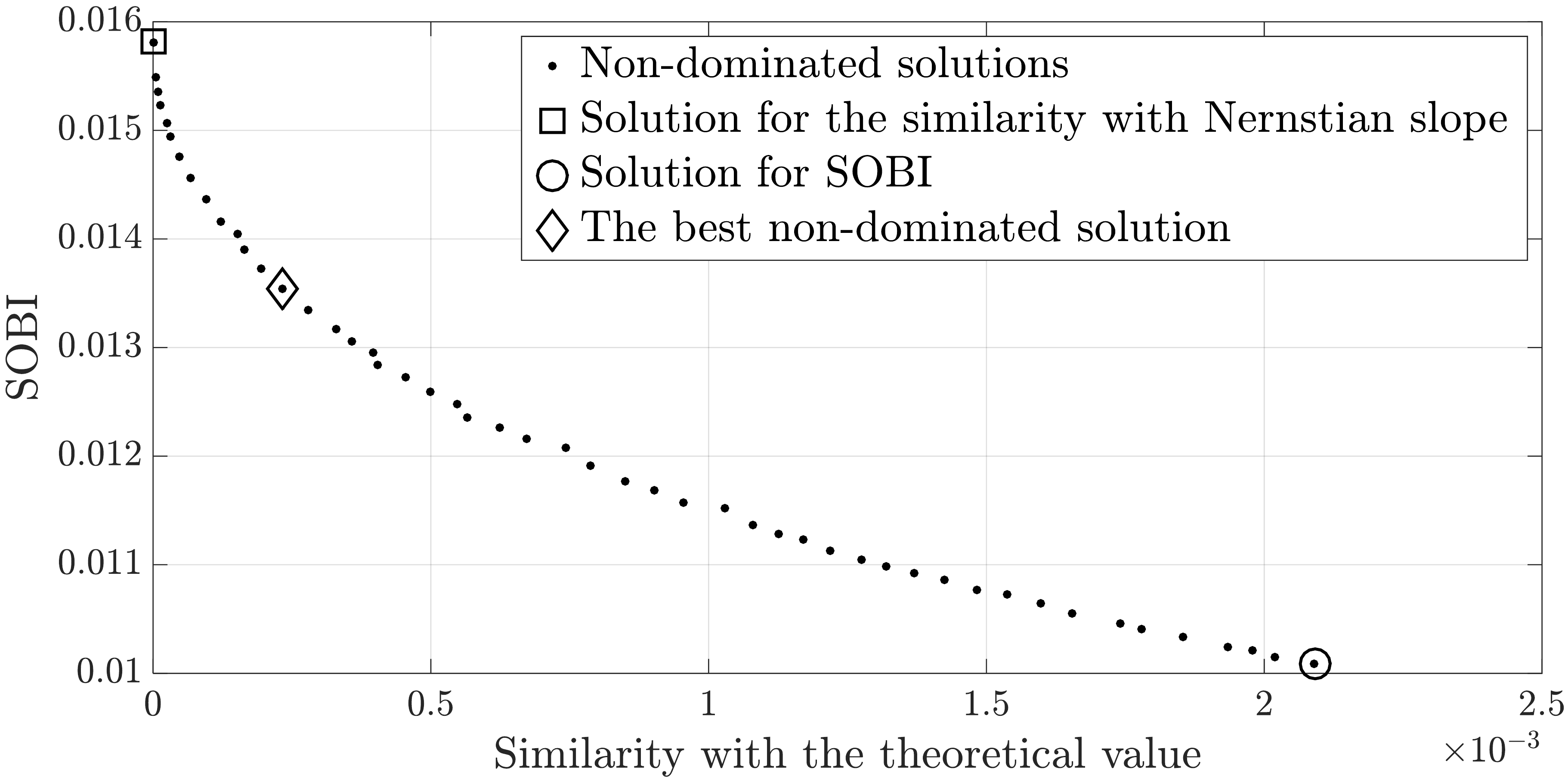}
\caption{Solutions obtained by the mono-objective and multi-objective approaches represented in the objective space composed by the similarity with the Nernstian slope and SOBI cost functions.}
\label{fig:paretoexp}
\end{figure}

The mixtures provided by the ISEs are shown in Figure~\ref{fig:mixt}. After applying the proposed BSS approach\footnote{The considered algorithm parameters was $L=100$, $\tilde{L}=50$, $\alpha=50$ and $G=30$. This parameters led to both convergence and diversity of the non-dominated set.}, which solves~\eqref{met:moo}, we obtained the set of non-dominated solutions illustrated in Figure~\ref{fig:paretoexp}, which also presents the solutions obtained minimizing both criteria individually (mono-objective approaches for similarity with Nernstian slope and SOBI-based cost function) and the best non-dominated one (the non-dominated solution that leads to the highest average SIR).

\begin{figure*}[ht]
\centering
\includegraphics[height=6.5cm]{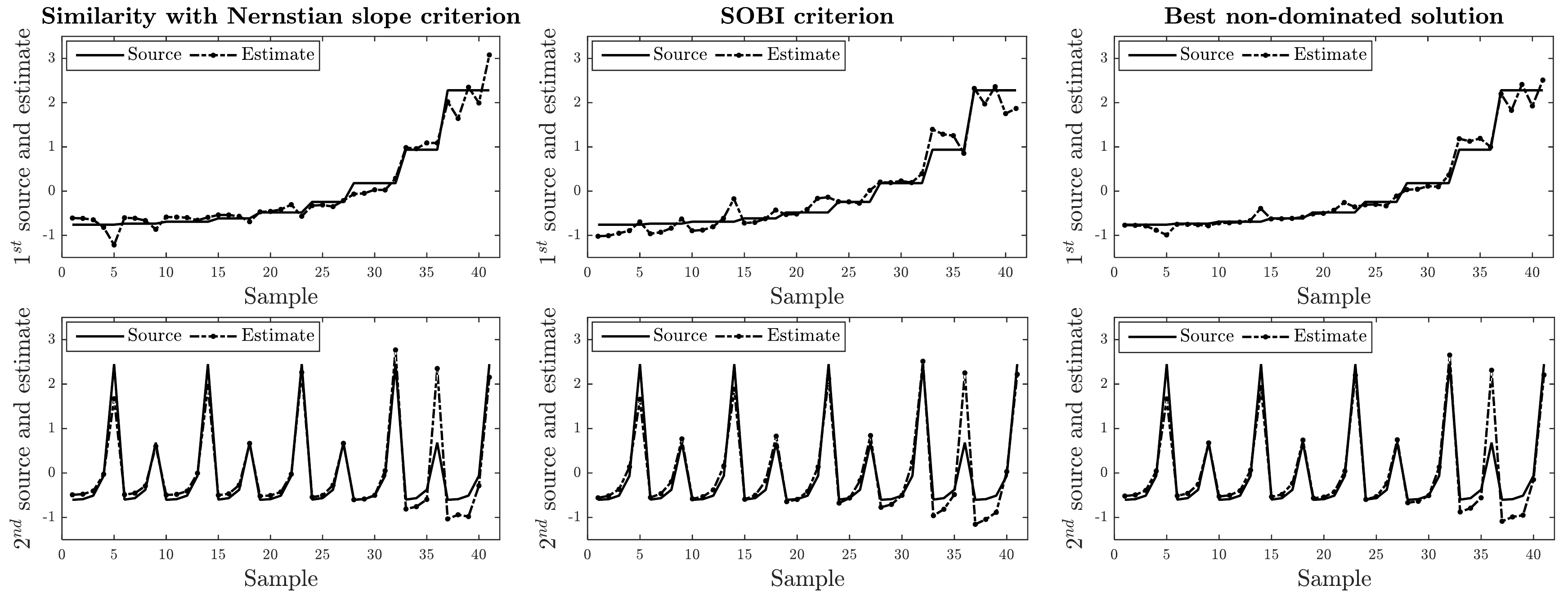}
\caption{Comparison between the sources and the estimates provided by the similarity with Nernstian slope criterion (mono-objective approach), SOBI criterion (mono-objective approach) and the best non-dominated solution (obtained by the proposed approach).}
\label{fig:estim}
\end{figure*}

The retrieved sources associated with the solutions highlighted in Figure~\ref{fig:paretoexp} are shown in Figure~\ref{fig:estim}. One can note that the best non-dominated solution obtained by the proposed multi-objective method is clearly less prone to the interference problem. Indeed, the obtained SIR values were given by SIR$_1=13.3$ dB and SIR$_2=9.3$ dB (for the mono-objective optimization of the cost function related to the similarity with Nerstian slope), SIR$_1=12.4$ dB and SIR$_2=9.2$ dB (for SOBI-based cost function) and SIR$_1=16.2$ dB and SIR$_2=9.3$ dB (the best non-dominated solution). 

We checked the computational time needed by each approach. For the mono-objective strategy based on the similarity with Nerstian slope, the required time was 0.0004 seconds. The mono-objective strategy based on the SOBI required 0.5400 seconds. Our proposal took 4.3200 seconds (30 iterations of SPEA2 algorithm, i.e. 0.1440 seconds each iteration).\footnote{Computing device: Intel Core i7, 2.20 GHz, 8.00 GB RAM, software MATLAB 2015.}

We also provide a comparison of the obtained results with existing works that consider a Bayesian approach~\cite{Duarte2014} (SIR$_1=4.4$ dB and SIR$_2=9.3$ dB), an ICA-based method~\cite{Duarte2014} (SIR$_1=11.0$ dB and SIR$_2=10.6$ dB) and a quadratic modeling~\cite{Ando2015} (SIR$_1=16.3$ dB and SIR$_2=11.1$ dB). Our proposal was able to provide comparable results to those obtained in~\cite{Ando2015}, which relies on a more complex mixing modeling than the one adopted in the present work. Such a feature is interesting since the solution~\cite{Ando2015} is prone to instability issues due to the use of recurrent separating systems.

%In Table~\ref{tab:comp_res}, we provide a comparison of the obtained results with those already found in the literature. Our proposal was able to provide comparable results to those obtained in~\cite{Ando2015}, which relies on a more complex mixing modeling than the one adopted in the present work. Such a feature is interesting since the solution~\cite{Ando2015} is prone to instability issues due to the use of recurrent separating systems.

%\begin{table}[htb!]
%  \begin{center}
%  \caption{Comparison of results.}\label{tab:comp_res}
%  {
%  \renewcommand{\arraystretch}{1.1}
%  \begin{tabular}{c c c }
%    \hline
%    \multirow{2}{*}{Approach} & \multicolumn{2}{c}{SIR (dB)} \\
%		\cline{2-3}
%     & Na$^+$ activity & K$^+$ activity \\
%    \hline
%    Bayesian method~\cite{Duarte2014} & 4.4 & 9.3 \\
		%\hline
%		ICA-based method~\cite{Duarte2014} & 11.0 & 10.6 \\
		%\hline
%    Quadratic modeling~\cite{Ando2015} & 16.3 & 11.1 \\
		%\hline
%    Multi-objective approach & 16.2 & 9.3 \\
%    \hline
%  \end{tabular}
%  }
%  \end{center}
%\end{table}

%\begin{figure}[ht]
%\centering 
%\includegraphics[height=6.3cm]{mixtures2.pdf}
%\caption{ISE responses.}
%\label{fig:mixt}
%\end{figure} 

\subsection{Selection of the best non-dominated solution}

\begin{figure}[ht]
\centering
\includegraphics[height=4.2cm]{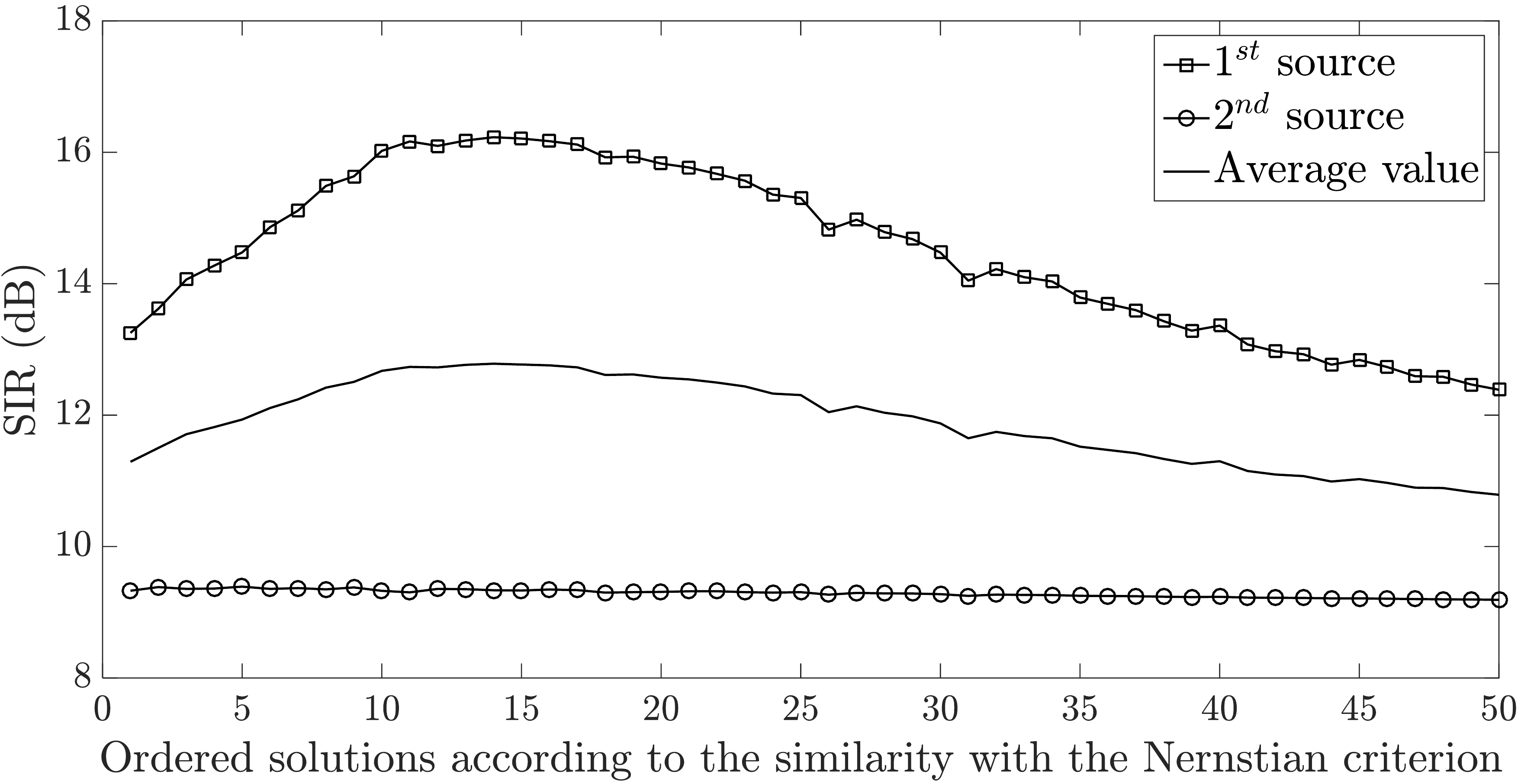}
\caption{SIR (dB) for all non-dominated solutions (these solutions are ordered according to the similarity with the Nernstian slope criterion).}
\label{fig:evolsir}
\end{figure}

Since the BSS problem is blind, it is not possible to select the best non-dominated solution according to SIR values. However, in practice, this selection could be done by incorporating subjective priors into the decision. For example, the user could vary the solutions within the non-dominated set in order to search for smooth solutions, which may lead to higher values of SIR\footnote{One may note that the best non-dominated solution is smoother than the ones provided by a mono-objective formulation.}. Moreover, as illustrated in Figure~\ref{fig:evolsir}, the SIR of the non-dominated solutions is very often higher than the SIR associated with the mono-objective optimization of the criteria expressed in~\eqref{met:moo}. Therefore, in the case under study, even a random choice of a non-dominated solution would perform better than the mono-objective solutions. 

\subsection{Robustness varying the Nernstian slope reference}

\begin{figure}[ht]
\centering
\includegraphics[height=4.1cm]{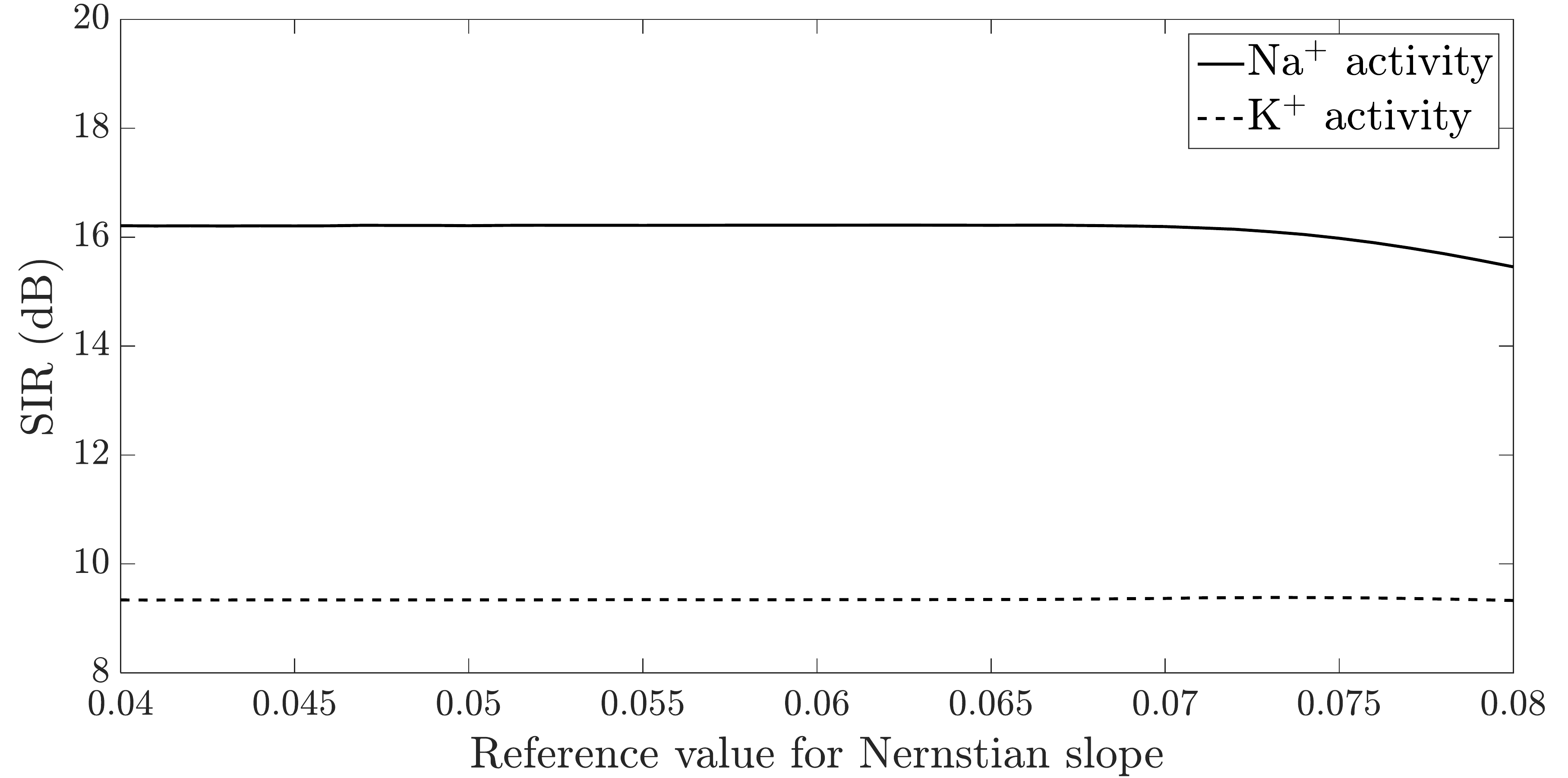}
\caption{SIR (dB) values for different reference values for the Nernstian slope.}
\label{fig:robust}
\end{figure}

Since in practice the reference value of $59$ mV may not be observed for the Nernstian slope, we checked the robustness of the proposed approach when different slopes are considered. We performed an experiment varying the reference value for $d_i^*$ in the range [40, 80] (mV). Figure~\ref{fig:robust} presents the best SIR values as a function of the reference values. Our proposal was able to achieve a good value of SIR even where there is a variation in Nernstian slope reference value; there is slight performance decrease for $d_i^*$ greater than 70 mV.

\section{Conclusion}
\label{sec:conclusion}

We introduced a multi-objective BSS approach to deal with the interference problem in ISEs. Experiments with an actual dataset pointed out that the proposal provides the user a set of non-dominated solutions which comprises the optimal solutions associated with the mono-objective optimization of the individual criteria. Moreover, the best non-dominated solution obtained in our experiments provided a better estimation of the sources when compared to mono-objective optimization. The computational time required by our proposal is greater than those required by mono-objective approaches but clearly acceptable in the context of chemical sensing. Finally, it is worth noticing that the proposed approach can be easily extended to take into account other criteria that are related to additional prior information on the mixing process and/or original sources. Future works will comprise analyses on the computational complexity of the proposed approach and the application on different datasets.

\bibliographystyle{IEEEtran}
\bibliography{IEEEabrv,_references}

\end{document}